\documentstyle[aps,preprint]{revtex}
\newcommand{\be}{\begin{equation}}
\newcommand{\ee}{\end{equation}}
\newcommand{\bea}{\begin{eqnarray}}
\newcommand{\eea}{\end{eqnarray}}

\def\6{\partial}

\def\a{{\alpha}}
\def\b{{\beta}} 
\draft
\begin{document}

\title{Bosonic D-Branes at Finite Temperature}
\author{I. V. Vancea\footnote{e-mail: ivancea@ift.unesp.br\\
On leave from Babes-Bolyai University of Cluj} }
\address{ Instituto de F\'{\i}sica Te\'{o}rica, Universidade 
Estadual Paulista (UNESP)\\
Rua Pamplona 145, 01405-900 S\~{a}o Paulo-SP, Brazil}

\maketitle


\begin{abstract}
We derive the finite temperature description of bosonic D-branes in the thermo
field approach. The results might be relevant to the study of thermical
properties of D-brane systems.
\end{abstract}


\vspace{1cm}

During the past several years, researches on the properties of D-branes,
viewed either as solitons of low energy field theories \cite{dkl} or as states
in perturbative spectra of strings \cite{pol}, has continued to grow unabated.
Theoretical studies predicted perturbative \cite{gpr} as well as
non-perturbative dualities \cite{npd} among string theories and M-theory.
D-branes offered new approaches to gauge theories \cite{ngt}, a first
microscopic description of the Beckenstein-Hawking entropy of black-holes
\cite{bhe} and new insights in string cosmology \cite{stc}. More recently,
D-branes have been used to conjecture a relationship between gravity and
quantum field theory \cite{mal} and to study the stable non-BPS states of
string spectra \cite{sen}. In lights of these achivements, its is worthwhile
to understand deeper the physical properties of D-branes.

In this Letter we aim at constructing bosonic D-brane at finite temperature as
boundary states of perturbative closed strings \cite{bs}. Although the
construction of a single D-brane works in principle for any temperature, in
considering statistical ensembles one should limit the discussion to low
temperatures mainly because of the following reason. The bosonic string
theories contain tachyons and, at zero mass level, dilatons and gravitons. The
presence of these fields make the ensemble of strings to suffer a first order
phase transition at $T_c < T_H$, where $T_H$ is the Hagedorn temperature.
As was argued in \cite{aw}, the latent heat involved in the transition is
large enough to break down the notion of temperature shortly above $T_c$ (for
a review see \cite{lag}.) The thermodynamics of a generic Dp-brane in the
imaginary time formalism was studied for the first time in \cite{mvm} while
the D-brane free energy  in external electro-magnetic field has been
computed in \cite{od} and in \cite{rsz} by using a boundary state formalism.  

In what follows, we are going to apply the thermo field theory \cite{tu} 
to construct the equations that define the D-branes and the boundary
states of closed strings that satisfy these equations. We note that the
quasi-particle picture in this theory no longer holds at sufficient large
temperatures. This might be another reason for studying the ensembles of
branes at low temperatures. However, it is not clear yet what is the validity
of the quasi-particle interpretation in the case of strings and to what extent
it may affect the ensembles of branes at high temperature. We hope to clarify
these aspects in a forthcoming paper \cite{cv}.

Consider a free bosonic closed string in the Minkowski spacetime and in the
conformal gauge at zero temperature $ T = 0$. We denote by $\a^{\mu }_{n}$ 
and $\b^{\mu}_{n}$, $\mu = 0,1,\ldots ,25$, $n \in Z^* $, the right- and 
left-moving modes, respectively. In order to apply the thermo field method we
pass to the oscillator description of string modes. This is given in terms of 
creation and annihilation operators for each mode $n$ and in each sector
\bea
A^{\mu \dagger}_{n} & = & \a^{\mu}_{-n}~~,~~A^{\mu}_{n} = \a^{\mu}_{n}
\nonumber\\
B^{\mu \dagger}_{n} & = & \b^{\mu}_{-n}~~,~~B^{\mu}_{n} = \b^{\mu}_{n},
\label{oscop}
\eea
where $n \in N^*$. In units such that $\hbar = 1$ the above operators describe 
independent harmonic oscillators of frequencies $\omega_n = n$. The vacuum
states of right- and left-moving sectors are denoted by $|0 \rangle_{\a}$ and
$|0 \rangle_{\b}$, respectively, and the fundamental state of string is given 
by the following product
\be
|0 \rangle = |0 \rangle_{\a} |0 \rangle_{\b} |p=0 \rangle ,
\label{vactzero}
\ee
where $|p \rangle$ is the eigenstate of the linear momentum of the 
center-of-mass corresponding to the eigenvalues $p^{\mu}$. 

A D$p$-brane is obtained by imposing the Neumann boundary conditions along
$a=0, 1, \ldots, p$ directions and the Dirichlet boundary conditions along 
$i=p+1, \ldots , 25$ directions of the open string at the endpoint $\sigma=0$.
These are equivalent with the following operatorial equations on the Hilbert
space of the closed string \cite{bs}
\bea
(A^{a}_{n} + B^{a \dagger}_{n})|B_{mat} \rangle & = &
(A^{a \dagger }_{n} + B^{a}_{n})|B_{mat} \rangle = 0
\nonumber\\     
(A^{i}_{n} - B^{i \dagger}_{n})|B_{mat} \rangle & = &
(A^{i \dagger }_{n} - B^{i}_{n})|B_{mat} \rangle = 0,
\label{brcond1}
\eea
for oscillator modes $ n > 0$, and
\be
{\hat{p}}^{a} |B_{mat} \rangle = ({\hat{X}}^{i} - y^i ) |B_{mat} \rangle = 0 ,
\label{brcond2}
\ee
for the momentum and position operators of the center-of-mass of string. Here,
$ \{ y^i \} $ represent the coordinates of the D$p$-brane in the transverse 
space. 

Let us construct the counterpart of equations (\ref{brcond1}) and 
(\ref{brcond2}) at $T \neq 0$ using the thermo field approach \cite{tu}. To 
this end, firstly we have to double the system. We denote by \~{} the 
quantities
that correspond to an {\em identical} copy of the original bosonic string. The
corresponding operators obey the usual commutation rules, the right- and
left-sector operators commute and they also commute with the operators 
corresponding to the original string. For example
\be
[ A^{\mu}_{n}, A^{\nu\dagger}_{m}] = [{\tilde{A}}^{\mu}_{n},
{\tilde{A}}^{\mu\dagger}_{m}] = \delta_{n,m}\eta^{\mu \nu}
\nonumber
\ee
\be
[A^{\mu}_{n},{\tilde{A}}^{\nu}_{m}] = [A^{\mu}_{n},{\tilde{A}}^{\nu\dagger}_{m}] 
= [A^{\mu}_{n},{\tilde{B}}^{\nu}_{m}]=\cdots = 0.
\label{comm}
\ee
The extended Hilbert space of the total system is given by the direct product 
of
the two Hilbert spaces of closed strings
\be
H_0 = H \times \tilde{H}
\label{hilbert}
\ee
and we denoted a state from $ H_0 $ by $|~~\rangle \rangle$. The vacuum state 
of the total system is given by
\bea
|0\rangle \rangle &=& |0 \rangle \rangle_{\a} |0 \rangle \rangle_{\b} 
= (|0\rangle_{\a} |\tilde{0}\rangle_{\a})
(|0\rangle_{\b}|\tilde{0}\rangle_{\b})
\nonumber\\
&=&(|0\rangle_{\a}|0\rangle_{\beta })
(|\tilde{0}\rangle_{\a}|\tilde{0}\rangle_{\beta }),
\label{vactzero1}
\eea
where the last equality is a consequence of the fact that the original string 
and the \~{} string are independent. The first line in (\ref{vactzero1}) shows
explicitely the doubling of each oscillator while the second one shows the
string-tilde string structure of the vacuum state. In order to obtain
the fundamental state of the enlarged system we have to multiply (\ref{vactzero1})
by $|p \rangle |\tilde{p} \rangle$. 

The thermal description of the system can be obtained from the above one by
acting with Bogoliubov operators $G^{\a}_{n}$ and $G^{\b}_{n}$ on each sector
of the Hilbert space and on the creation and annihilation operators. 
The $G$-operators are defined as follows
\bea
G^{\a}_{n} &=& -i\theta (\b_T)(A_{n}\cdot{\tilde{A}}_{n} - 
A^{\dagger}_{n}\cdot{\tilde{A}}^{\dagger}_{n})
\nonumber\\
G^{\b}_{n} &=& -i\theta (\b_T)(B_{n}\cdot{\tilde{B}}_{n} - 
B^{\dagger}_{n}\cdot{\tilde{B}}^{\dagger}_{n}) .
\label{gop}
\eea
Here, $\theta_n(\b_T)$ is real and depends on the statistics of the $n$th 
oscillator \cite{tu}
\be
\cosh \theta_n(\b_T) = (1-e^{-\b_T n})^{-\frac{1}{2}}.
\label{theta}
\ee
Therefore, $\theta$ is the same in both right- and left-sectors for a given
mode $n$. The dot in (\ref{gop}) means the usual scalar product in the 
Minkowski space $A_n \cdot {\tilde{A}}_n = A^{\mu}_n \cdot {\tilde{A}}_{n\mu}$.
Since the oscillators are independent, the vacuum state at $T\neq 0$ has the 
following structure
\be
|0(\b_T ) \rangle \rangle = \prod_{n>0}e^{-iG^{\a}_{n}}
|0\rangle\rangle_{\a}
\prod_{m>0}e^{-iG^{\b}_{m}}|0\rangle\rangle_{\b}.
\label{vactnzero}
\ee
The creation and annihilation operators at $T \neq 0$ corresponding to 
(\ref{vactnzero}) are obtained by acting on the zero temperature operators
$ \{ A^{\dagger}, A, {\tilde{A}}^{\dagger}, \tilde{A} \}$ and 
$ \{ B^{\dagger}, B, {\tilde{B}}^{\dagger}, \tilde{B} \}$ with the 
$G$-operators
given in (\ref{gop}). Taking into account the algebraic properties of 
$G^{\a}_{n}$ and $G^{\b}_{n}$ it is easy to show that the finite temperature 
annihilation operators can be cast into the following form
\bea
A^{\mu}_{n} (\b_T) & = & e^{-iG^{\a}_{n}}A^{\mu}_{n}e^{iG^{\a}_{n}}=
u_n(\b_T)A^{\mu}_{n} - v_{n}(\b_T){\tilde{A}}^{\mu\dagger}_{n}
\nonumber\\
{\tilde{A}}^{\mu}_{n} (\b_T) & = & e^{-iG^{\a}_{n}}{\tilde{A}}^{\mu}_{n}
e^{iG^{\a}_{n}}=
u_n(\b_T){\tilde{A}}^{\mu}_{n} - v_{n}(\b_T)A^{\mu\dagger}_{n}
\label{trbogop1}
\eea
in the right-moving sector and 
\bea
B^{\mu}_{n} (\b_T) & = & e^{-iG^{\b}_{n}}B^{\mu}_{n}e^{iG^{\b}_{n}}=
u_n(\b_T)B^{\mu}_{n} - v_{n}(\b_T){\tilde{B}}^{\mu\dagger}_{n}
\nonumber\\
{\tilde{B}}^{\mu}_{n} (\b_T) & = & e^{-iG^{\b}_{n}}{\tilde{B}}^{\mu}_{n}
e^{iG^{\b}_{n}}=
u_n(\b_T){\tilde{B}}^{\mu}_{n} - v_{n}(\b_T)B^{\mu\dagger}_{n}
\label{trbogop2}
\eea
in the left-moving sector. Here, 
\be
u_n(\b_T)=\cosh \theta_n (\b_T)~~,~~
v_n(\b_T)=\sinh \theta_n (\b_T).
\label{uv}
\ee
Since the operators $\hat{p}$, $\hat{X}$, $\hat{\tilde{p}}$ and 
$\hat{\tilde{X}}$ commute with all oscillator operators, they are not 
affected
by the transformations of the type (\ref{trbogop1}) and (\ref{trbogop2}). If
we construct zero mode $G$-operators as we did for the oscillators
we see that the momenta commute with them. Therefore, we take the position 
and momenta operators of both the string and the tilde string to be invariant
under the transformations above. The corresponding eigenstates of the momenta
operators are taken to be invariant, too. Then it is not difficult 
to 
see that the string coordinate operators $X^{\mu}(\tau , \sigma )(\b_T)$ 
and
${\tilde{X}}^{\mu}(\tau , \sigma )(\b_T)$ constructed from (\ref{trbogop1})
and (\ref{trbogop2}) are solutions of the string equations of motion 
($ G^{\a}_{n} $ and $ G^{\b}_{n} $ do not act on the two dimensional wave 
functions.) Therefore, we may construct the D$p$-brane states at $ T\neq 0 $ 
by
imposing the Neumann and Dirichlet boundary conditions on the appropriate
spacetime directions
\be
\6_{\tau}X^{a}(\tau , \sigma )(\b_T )|_{\tau =0 } = 0~~,~~
X^{i}(\tau , \sigma )(\b_T )|_{\tau =0 } = y^i ,
\label{bct}
\ee
and similarly for ${\tilde{X}}^{\mu}(\tau , \sigma )(\b_T)$. Here, 
$a=0,1,\ldots,p$ and $i=p+1,\ldots,25$. When imposed on the extended Hilbert
space, these relations define two sets of equations that determine D$p$- and
\~{D}$p$-boundary states at $T\neq 0$. We denote these states
by $|B_{mat}(\b_T )\rangle\rangle$ and
$|{\tilde{B}}_{mat}(\b_T)\rangle \rangle$,
respectively. If we introduce the folowing matrices \cite{bs}
\be
S^{\mu \nu} = (\eta^{ab},-\delta^{ij})~~,~~{\tilde{S}}^{\mu \nu} =
({\tilde{\eta}}^{ab},-{\tilde{\delta}}^{ij}),
\label{matr}
\ee
we can see that the equations defining D$p$- and \~{D}$p$-branes have the
following form
\be
[u_{n}(\b_T)(A^{\mu}_{n} + S^{\mu}_{\nu}B^{\nu\dagger}_{n}) +
v_{n}(\b_T)({\tilde{B}}^{\mu}_n +
{\tilde{S}}^{\mu}_{\nu}{\tilde{A}}^{\nu\dagger}_n)]|B_{m}(\b_T)\rangle\rangle
=0 
\label{dpb1}
\ee
\be
[u_{n}(\b_T)(A^{\mu\dagger}_{n} + S^{\mu}_{\nu}B^{\nu}_{n}) +
v_{n}(\b_T)({\tilde{B}}^{\mu\dagger}_n +
{\tilde{S}}^{\mu}_{\nu}{\tilde{A}}^{\nu}_n)]|B_{m}(\b_T)\rangle\rangle
=0
\label{dpb2}
\ee
and
\be
[u_{n}(\b_T)({\tilde{A}}^{\mu}_{n} +
{\tilde{S}}^{\mu}_{\nu}{\tilde{B}}^{\nu\dagger}_{n}) +
v_{n}(\b_T)(B^{\mu}_n +
S^{\mu}_{\nu}A^{\nu\dagger}_n)]|{\tilde{B}}_{m}(\b_T)\rangle\rangle
=0 \label{tdpb1}
\ee
\be
[u_{n}(\b_T)({\tilde{A}}^{\mu\dagger}_{n} +
{\tilde{S}}^{\mu}_{\nu}{\tilde{B}}^{\nu}_{n}) +
v_{n}(\b_T)(B^{\mu\dagger}_n +
S^{\mu}_{\nu}A^{\nu\dagger}_n)]|{\tilde{B}}_{m}(\b_T)\rangle\rangle
=0 
\label{tdpb2}
\ee
If for some mode $n$ and some critical temperature $T_c$
the equation $ u_n(\b_{T_c})= v_n(\b_{T_c})$ holds,
then the corresponding equations (\ref{dpb1}),(\ref{dpb2})
and (\ref{tdpb1}),(\ref{tdpb2}) are identical and give the same $n$th
oscillator contribution to $|B_{mat}(\b_T )\rangle\rangle$ and
$|{\tilde{B}}_{mat}(\b_T)\rangle \rangle$. This condition is actually
equivalent to the limit $T_c \rightarrow \infty$ for any finite $n$, case in
which (\ref{dpb1}),(\ref{dpb2}) and (\ref{tdpb1}),(\ref{tdpb2}) are identical
for all $n$. The equations above are consistent with the zero temperature
description of the D$p$-branes as boundary states since in the limit
$T \rightarrow 0$ they reduce to two copies of the equations (\ref{brcond1}).

Let us look for the solutions of the equations above. We put the state
$|B_{mat}(\b_T )\rangle\rangle$ under the form
\be
|B_{mat}(\b_T )\rangle\rangle = {\hat{B}}_{mat}|0(\b_T)\rangle\rangle
={\hat{B}}_{mat}(\b_T)|0\rangle\rangle .
\label{bsttt}
\ee
Then, by analogy with the zero temperature limit, we can write the operator
${\hat{B}}_{mat}(\b_T)$ in the following way
\be
{\hat{B}}_{mat}(\b_T) \sim \delta^{25-p}({\hat{X}}^{i}-y^i){\hat{O}}_{mat}
\prod_{k=1}^{\infty}e^{iG_k}.
\label{opo}
\ee
After some simple algebra one can see that the equations
(\ref{dpb1}),(\ref{dpb2}) have solution of the form (\ref{bsttt}). Moreover,
this solution is not unique since it can be constructed with the following
different operators
\bea
{\hat{O}}_{mat}^1 & = & \prod_{n=1}^{\infty}e^{-A^{\dagger}_n S B^{\dagger}_n}
\prod_{r=1}^{\infty}e^{-
{\tilde{B}}^{\dagger}_r \tilde{S} {\tilde{A}}^{\dagger}_r}
\nonumber\\
{\hat{O}}_{mat}^{2,3} & = & \prod_{n=1}^{\infty}e^{-A^{\dagger}_n S 
B^{\dagger}_n}
\times 1 
\pm 1 \times
\prod_{n=1}^{\infty}e^{-
{\tilde{B}}^{\dagger}_n \tilde{S} {\tilde{A}}^{\dagger}_n}
\label{opoo}.
\eea
Thus, the D$p$-brane states at $T \neq 0 $ from the extended Hilbert space
are degenerate in the sense of the equations above. The same considerations
apply to $ |{\tilde{B}}_{mat}(\b_T)\rangle \rangle $ states. In this case we
obtain the first solutions from (\ref{opoo}) as a consequence of the fact that
the right- and left-modes commute for either the original string as well as
for its copy. Each of D$p$- or
\~{D}$p$-branes depend on the oscillator operators of the original
as well as the tilde strings. However, according to the thermofield theory,
the physical quantities in a thermal D$p$-brane state may depend only on the
vev of the operators without tilde \cite{tu}. In this sense, the solutions
derived above in (\ref{opoo}) are equivalent. Note that one has to include
in $|B_{mat}(\b_T )\rangle\rangle$ the eigenvectors of momenta operators at 
$p=0$. However, since these operators and their eigenvalues do not
transform under $G$-transofrmations, their contribution is the same as in $T=0$
case. The normalization constant
that enter (\ref{bsttt}) is related, at $T=0$, to the tension of the brane
\cite{pol}. To determine it at finite temperature, one has to perform
calculations in both open and closed string sectors \cite{cv}.

The above boundary states make sense only if they belong to the physical sector
of the theory. At $T=0$, this sector is defined using the conformal invariance 
of the string theory. The same definition can be applied here since the 
conformal invariance is not broken. Indeed, using the fact that $G$-operators
commute, it is easy to show that the string oscillation modes statisfy the same
algebra at finite temperature. Then, since 
${X}^{\mu}(\tau , \sigma )(\b_T)$ and
${\tilde{X}}^{\mu}(\tau , \sigma )(\b_T)$
satisfy the string equations of motion
the operators $L_n (\b_T )$ and
${\tilde{L}}_n (\b_T )$
have formally the same expansion 
in terms of modes $\a_m (\b_T )$ as the Virasoro operators at zero temperature
and satisfy the same algebra in both right- and left-sectors and for both the
original string and the tilde string. The algebras in all of these sectors are 
independent of each other. Using the conformal symmetry, one can impose the 
physical state conditions on the Fock space of the extended system as for $T=0$
\cite{cv}. Also, the problem of mantaining the conformal invariance 
is related to the possibility of mapping the open string sector to the 
closed one and to constructing the BRST invariant states. Note that the 
conformal invariance at finite temperature is due to the specific form of 
$G$-operators taken in (\ref{gop}). This is not the most general form allowed
in thermo field approach \cite{tu}. Nevertheless, the
conformal symmetry is mantained by other tranformations, too. The point here is
that $G$-operators do not mix the various oscillation mode which is a 
reasonable assumption at low temperature. At high temperature we might expect 
that 
$G$-operators involve interacting terms between various oscillators, case in 
which the conformal invariance is not automatically satisfied. If we stick on
the definition of strings as conformal invariant field theories in two 
dimensions, the conformal invariance imposes some constraints on 
$G$-operators\cite{cv}. On the other hand, breaking the conformal invariance by
$G$-operators might be a consequence of the fact that the temperature is not 
well defined for string systems at $T>T_c$ as was showed in \cite{aw}. 

In the case of super D-branes one has to repeat the same construction in
both NS-NS and R-R sectors of the closed string. It is quite easy to 
construct the corresponding Bogoliubov transformations for NS and R spinors
and the corresponding fermionic modes at finite temperature. However, it is
not clear yet what is the natural extension of the supersymmetry at 
$T \neq 0$ of the theory that is supersymmetric at $T = 0$ and whether the
superconformal invariance is broken or not \cite{cv}. The supersymmetry
in right- and left-modes splits into
\bea
\a ' p^{\mu} + \sum_n u_n(\b_T)\6_{\mp}X^{\mu}_n
\stackrel{susy}{\sim} \sum_{t}u_t(\b_T)\psi^{\mu}_{\mp (t)}
\nonumber\\
\sum_n v_n(\b_T)\6_{\mp}{\tilde{Y}}^{\mu}_n
\stackrel{susy}{\sim} - \sum_{t}v_t(\b_T){\tilde{\chi}}^{\mu}_{\mp(t)},
\label{susy}
\eea
where by ${\tilde{Y}}^{\mu}_n$ and ${\tilde{\chi}}^{\mu}_n$ we denoted the
bosonic and fermionic modes of the tilde string present in the equations of
the original string. (Similar expressions should be considered for the tilde
string, too.) The problem is
due to the presence  of $u_n(\b_T)$ and $v_n(\b_T)$ 
in the bosonic mode expansion
and to similar terms $u_t(\b_T)$ and $v_t(\b_T)$ in the fermionic mode 
expansion, where $t \in Z^+ + 1/2$ in NS sector and $t \in Z^+$ in R sector,
respectively. We note that the second set of coefficients is different from
the first one since it is related to Fermi-Dirac distributions of the 
fermionic modes. 

In summary, we have used the thermo field theory to find out the equations
that define the bosonic D$p$-branes at finite temperature. We found that the
corresponding boundary states from the extended Hilbert space are degenerate,
but equivalent at thermical equilibrium. They do not come in pairs since the
solutions are the same for the equations of the original string and for the
tilde string. In this respect the states that describe D$p$-branes and 
\~{D}$p$-branes are degenerate, opposite to the quasi-particle states in 
field theory. 

\acknowledgements

I have greatly benefited from discussion with M. C. B. Abdalla, N. Berkovits,
A. L. Gadelha, B. M. Pimentel, H. Q. Placido, D. L. Nedel and B. C. Vallillo. 
I also
acknowledge to M. A. De Andrade and J. A. Helayel-Neto for hospitality at
GFT-UCP where part of this work was done and to A. Sen for correspondence. The
work was financially supported by a FAPESP postdoc fellowship.

\end{document}